\documentclass{article}
\usepackage{graphicx}
\usepackage{geometry}
\begin{document}

\title{Nonequilibrium phase transition of a one dimensional system reaches the absorbing state by two different ways} 
\author{M. Ali Saif\\
Department of Physics,\\
University of Amran,
Amran,Yemen\\
masali73@gmail.com}

\maketitle
\begin{abstract}
We study the nonequilibrium phase transitions from the absorbing phase to the active phase for the model of disease spreading (Susceptible-Infected-Refractory-Susceptible (SIRS)) on a regular one dimensional lattice. In this model, particles of three species (S, I and R) on a lattice react as follows: $S+I\rightarrow 2I$ with probability $\lambda$, $I\rightarrow R$ after infection time $\tau_I$ and $R\rightarrow I$ after recovery time $\tau_R$. In the case of $\tau_R>\tau_I$, this model has been found to has two critical thresholds separate the active phase from absorbing phases \cite{ali1}. The first critical threshold $\lambda_{c1}$ is corresponding to a low infection probability and second critical threshold $\lambda_{c2}$ is corresponding to a high infection probability. At the first critical threshold $\lambda_{c1}$, our Monte Carlo simulations of this model suggest the phase transition to be of directed percolation class (DP). However, at the second critical threshold $\lambda_{c2}$ we observe that, the system becomes so sensitive to initial values conditions which suggests the phase transition to be discontinuous transition. We confirm this result using order parameter quasistationary probability distribution and finite-size analysis for this model at $\lambda_{c2}$. Additionally, the typical space-time evolution of this model at $\lambda_{c2}$ shows that, the spreading of active particles are compact in a behavior which remind us the spreading behavior in the compact directed percolation.

\end{abstract}

\section{Introduction}
Nonequilibrium phase transition from the active states to the absorbing states has been attracted a lot of scientific community efforts recently \cite{hen,hin,gez}. One of the most important efforts in this field is concerned by classify nonequilibrium systems into universal classes. In this sense, the directed percolation class (DP) is the most important class in the nonequilibrium phase transition to the absorbing states. Many models have been found to be belongings to this class, for example a contact process (CP), Domany-Kinzel cellular automaton (DK), Ziff-Gulari-Barshad (ZGB) model, pair-contact process (PCP), threshold transfer process (TTP) and branching annihilating walks with an odd number of offspring \cite{zif,voi,fer,fer1,zif1,tak,hoe}. According to Janssen-Grassberger criterion \cite{hin,gez,mar0} a model should belong to DP universality class if the model satisfies the following conditions, display a continuous transition into a unique absorbing state with a positive one-component order parameter, with short-range interactions and without quenched disorder or additional symmetries. In fact DP class seems to be even more general and has been found that, some systems belong to this universality class even if they are violate some of previous criterion for example in the long-range interactions \cite{she} or certain models with many absorbing state \cite{alb,mun1,mun2,mun3} or fluctuating passive states \cite{men}. 

Another important class of the nonequilibrium phase transitions to absorbing state is the voter universality class. This class has been observed in special case of models with a two symmetric ($Z_2$ symmetry) absorbing state \cite{gez,dor1,ham,dor2}. Models such as compact directed percolation (CDP), the $2A\rightarrow \phi$ and the $2A\rightarrow A$, the cellular automaton version of the nonequilibrium kinetic Ising model and L$\acute{e}$vy-flight anomalous diffusion in annihilating random walks belong to this class. Parity-Conserving universality class (PC) \cite{gez,dor1,ham,dor2} is another universality class to absorbing state. This class characterizes those models which conserve the number of particles modulo $2$. Examples are the one-dimensional kinetic Ising models which combined finite temperature spin exchange dynamics and zero temperature spin-flip \cite{me}, branching and annihilating random walks with even number of offspring \cite{zho} and parity-conserving class of surface-catalytic models \cite{zhu}.

Nonequilibrium discontinuous phase transitions from an active state to an absorbing state have been also observed in the dimensions higher than one in many cases. For example in a two dimensional ZGB model and its modifications \cite{zif,fer,fer1,cha,tom,eva,oli3,alb1}, a two dimensional reaction-diffusion contact-process-like model \cite{win,mar,dic}, a two lattice versions of the second Schl$\ddot{o}$gl model (SSM) \cite{oli1,oli2}, a two dimensional naming game model \cite{bar,bri,net}, two and four dimensions deterministic conserved threshold transfer process \cite{lee1,lee2} and the prisoner's dilemma with semi-synchronous updates \cite{ali} on two dimension. However, discontinuous phase transition to absorbing states have been rarely seen in one dimension. This is due to the fact that, the fluctuations in low dimensions are strong which make the continuous phase transitions are likely to occur. Hinrichsen \cite{hen,hin} argued that, first order phase transition can not occur in one dimensional nonequilibrium systems unless there are extra symmetries, conservation laws, long-range interactions or special boundary conditions. By any means in a one dimension, the first order phase transition has been observed in systems with conserved density \cite{lee2,meny}, models with long-range interactions \cite{odo} and in the systems with multi-component \cite{god,eva1}. For a two-species reaction-diffusion process on a one dimension the renormalization group methods predict a first order phase transition \cite{oer,wij}, however the numerical simulations of that model have been yielded results in disagreement with the renormalization group prediction \cite{dic1,mai}. The model candidate to violate Hinrichsen argumentations is the triplet creation model (TCM). This model is a single component and does not possess a conservation law or long-range interactions. Preceding study by Dickman and Tom$\acute{e}$ \cite{dic} had been suggested the first order phase transition of this model for a high value of diffusion rate ($D\geq 0.9$). In sequence, Cardoso and Fontanari modified that value to be $D\geq 0.95$ \cite{car}. Recently, the simulations results of TCM model by Park \cite{par1} have been shown that, the phase transition of this model is continuous for any value of $D\leq 0.98$. More recent study of this model by $\acute{O}$dor and Dickman \cite{gez2} suggests a continuous phase transition for any value of $D<1$.

In this work we are going to study the phase transition from the absorbing state to the active state of the epidemic spreading model SIRS (Susceptible- Infected- Refractory- Susceptible) on a one dimensional regular network. This model has been proven to has a two critical threshold \cite{ali1}. We are interested to study the phase transition close to those critical thresholds. This work is organized as follows. In section 2, the model and simulations methods are described. Simulation results close to the first critical point of this model are presented and discussed in section 3. Simulation results close to the second critical point of this model are given and discussed in section 4. Conclusions are given in section 5. 

\section{Model and Methods}
The model of epidemic spreading SIRS on the networks, can be described as follows \cite{kup}: consider a population of $N$ particles residing on the sites of a lattice in which each particle is connected to $k$ of its neighbors. The particles can exist in one stage of three stages, susceptible $(S)$, infected $(I)$ and refractory $(R)$. The interaction between the particles on the lattice is as follows: the particles in state $I$ on the network can infect any one of their neighbors which are in state $S$ with probability $\lambda$ at each time step ($S+I\rightarrow 2I$). The particles in state $I$ pass to the $R$ state after an infection time $\tau_I$ ($I\rightarrow R$). The particles in state $R$ return to the $S$ state after a recovery time $\tau_R$ ($R\rightarrow I$). During the $R$ phase, the particles are immune and do not infect.

For this model on the networks as it have been proven in Ref. \cite{ali1}, we have to distinguish between the following two cases: The first case happens when $\tau_I\geq \tau_R$, where in this case SIRS model has only a one critical threshold $\lambda_{c}$ separates the active phase from absorbing phase. Second case happens when $\tau_I\leq \tau_R$, in this case the SIRS model has a two critical threshold $\lambda_{c1}$ and $\lambda_{c2}$ in which the system is active in between them and die outside of them. In the second case, the first critical threshold $\lambda_{c1}$ is corresponding to the situation where the infection probability $\lambda$ is low. So in this case, the spreading of infection is limited and local therefore and when $\lambda<\lambda_{c1}$, the system will evolve to the absorbing state where all particles become susceptible (state $S$). In contrast the second critical threshold $\lambda_{c2}$ corresponds to the situation where the infection probability $\lambda$ is high. Therefore, in this case the infection will spread globally and quickly through the entire network. Now let us ask this question: What will happen for the system when $\lambda$ is high enough such that, all the particles in the network become infected during a time which is less than or equal to $\tau_I$?. In this case where $\tau_R>\tau_I$, so all the particles will approach the state $R$ followed by $S$ state during a time which is not longer than $\tau_R$. Then this case is also an absorbing state for this model. However this absorbing state is un-stationary where the particles will stay in this state only for a time which is not longer than $\tau_R$ after that the system will approach the stable absorbing state $S$ (see Fig. 1). Hence we can say that, when $\tau_I\leq \tau_R$ the SIRS model has a two absorbing states, even the second absorbing state is not stable but if the system reaches it, the system will evolve surely to the stable absorbing state (first absorbing state $S$). 

As aforementioned, the absorbing state of this model is the state where the lattice becomes free of infected particles, i. e. the state $S$. SIRS model approaches this absorbing state by two different ways. At $\lambda_{c1}$ the model reaches absorbing state due to of that, the strength of infection is very low hence, the average number of susceptible particles infected by an already infected one during the time $\tau_I$ is less than one. Whereas at $\lambda_{c2}$ the strength of infection is high such that, each infected particle infects all of its neighbors during the time $\tau_I$. Second critical threshold is equivalent to the state where all particles reach the state $I$ during a time which is less than or equal $\tau_I$. In this case and where $\tau_R>\tau_I$ then, instantaneously all the particles will approach the state $R$ followed by the absorbing state $S$ during a time which is not longer than $\tau_R$. we can consider the state where all particles on the lattice are infected (state $I$) as an absorbing state of this model however, this state is un-stationary will end up to the absorbing state $S$.
In Fig. 1 we show the space-time evolution for a one dimensional lattice of $11$ sites with periodic boundary condition. In this lattice each particle is connected to its first two neighbors. We set the infection probability $\lambda=1$, infection time $\tau_I=2$, recovery time $\tau_R=3$ and the infection starts with one particle on the center of lattice. It is clear from the figure that, all particles on lattice will approach the absorbing state $S$ after $11$ time steps.

\begin{figure}[htb]
 \includegraphics[width=50mm,height=50mm]{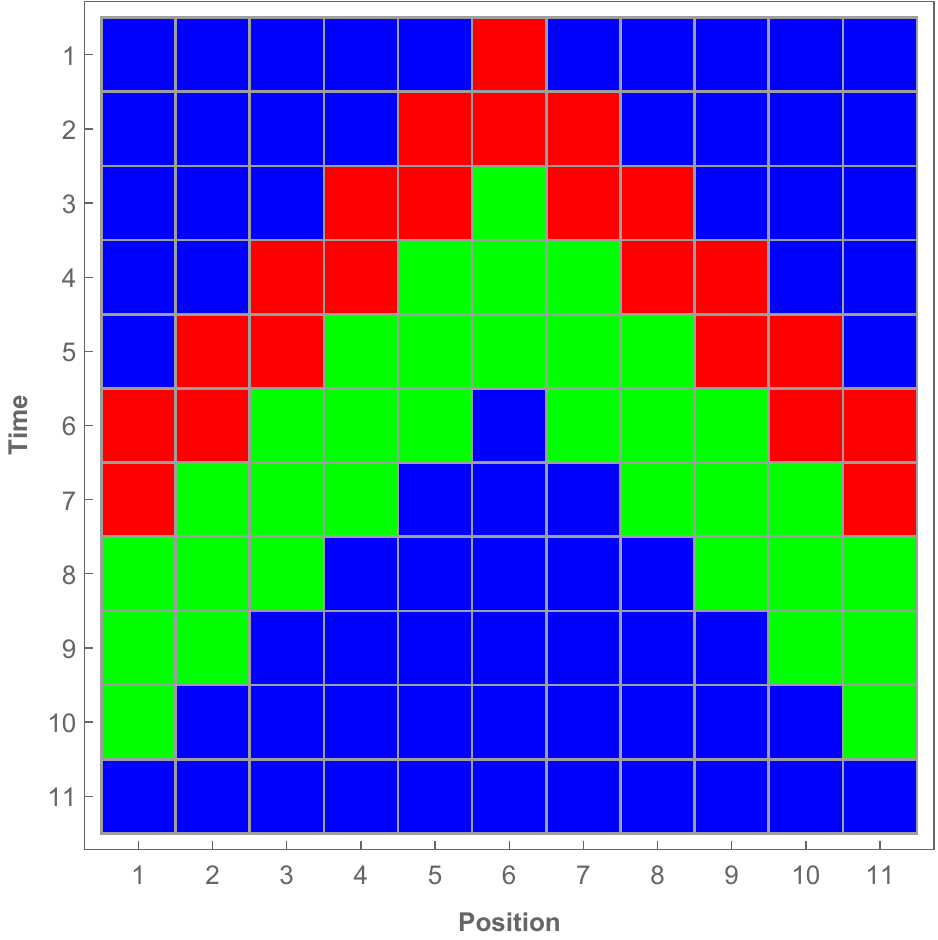}
\caption{Space-time evolution of lattice of $11$ sites with periodic boundary condition when $\lambda=1$, $\tau_I=2$, $\tau_R=3$ and $k=1$ (Blue: $S$, Red: $I$ and Green: $R$).}
 \end{figure}

We simulate this model on a regular one dimensional lattice with periodic boundary conditions in which each particle on the lattice is connected to $k=3$ of its nearest neighbors on each side. The system updates synchronically. In this work, we fix the values of infection time and recovery time to be $\tau_I=7$ and $\tau_R=9$ unless we state different. The order parameter $\rho(t)$ is the density of infective particles $I$ (active particles) 
\begin{eqnarray}
\rho(t)=\frac{\left\langle \sum_j I_j(t)\right\rangle}{N}
\end{eqnarray}
 where $N$ is the total number of lattice sites and $\left\langle ...\right\rangle$ stands for average over ensembles. Steady state of order parameter $\rho_s$ is the state when $\rho_s \equiv{\stackrel{lim}{t\rightarrow \infty}}\rho(t)$.
  
In Fig. 2, we recreate the steady state of the density of active particle as function of $\lambda$ given in Ref. \cite{ali1}. For each point in Fig. 2, we start the simulation from the initial density of active particle $\rho(0)=0.1$ and averaged over $100$ configurations after discarding $10^4$ initial time steps. Figure shows clearly the two critical thresholds which we are interesting to study the phase transition at them.
\begin{figure}[htb]
 \includegraphics[width=100mm,height=80mm]{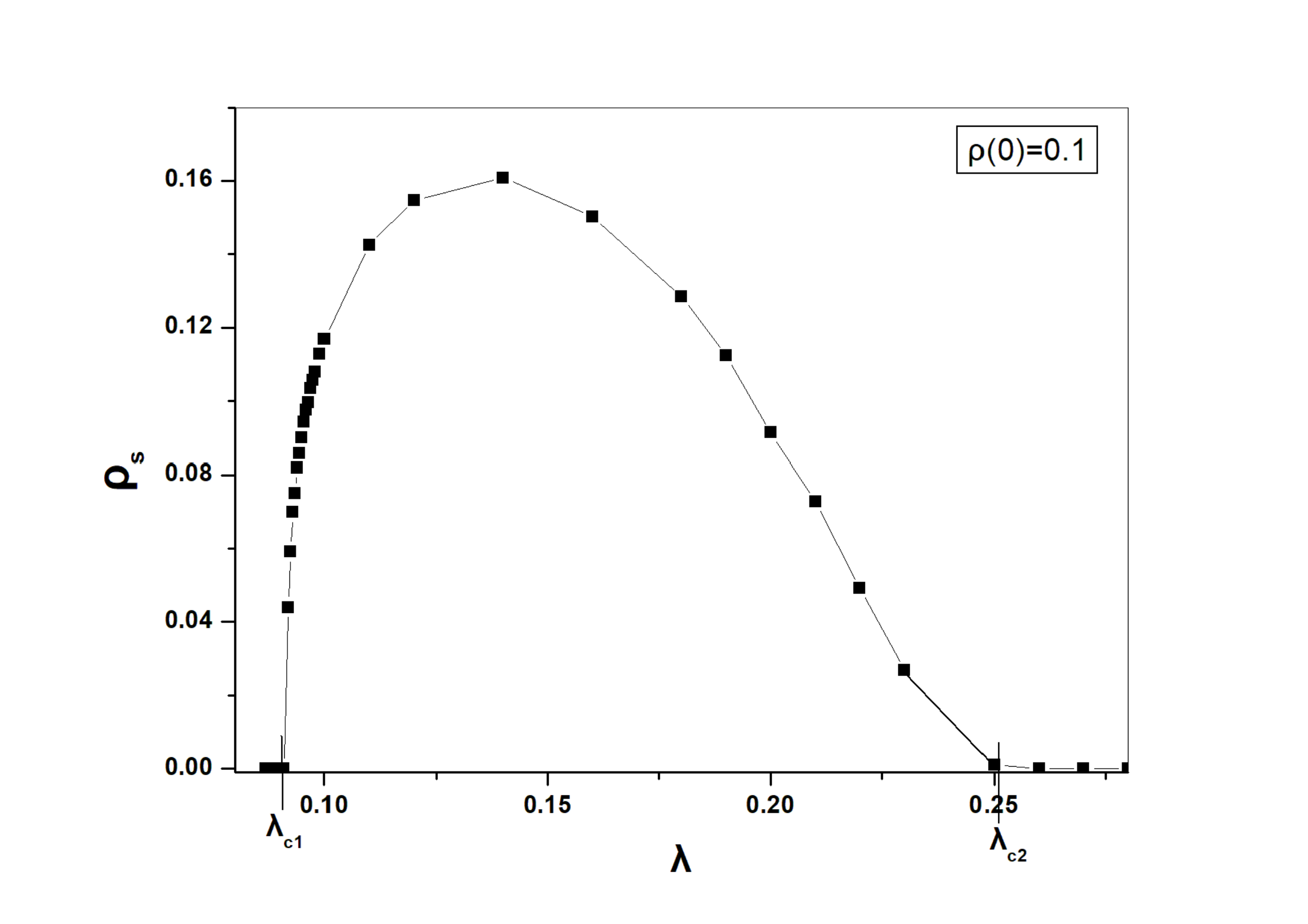}
\caption{The steady state of the density of particles $\rho_s$ at various values of the $\lambda$ when $N=10^4$, $k=3$, $\tau_I=7$ and $\tau_R=9$.}
 \end{figure} 

\section{Phase Translation at the first critical point $\lambda_{c1}$}      
For a general view about the kind of the phase transition at the first critical point, we start the simulation of this model with the typical space-time evolution beside the $\lambda_{c1}$. In Fig. 3 we show the typical space-time evolution of this model when the simulation starts initially from a single active seed located at the center of lattice for the values of the parameters $\lambda=0.090$ and $\lambda=0.094$. Fig. 3 seems to be similar to the typical space-time evolution of the systems which undergo DP phase transition \cite{hin}. 
\begin{figure}[htb]
 \includegraphics[width=50mm,height=50mm]{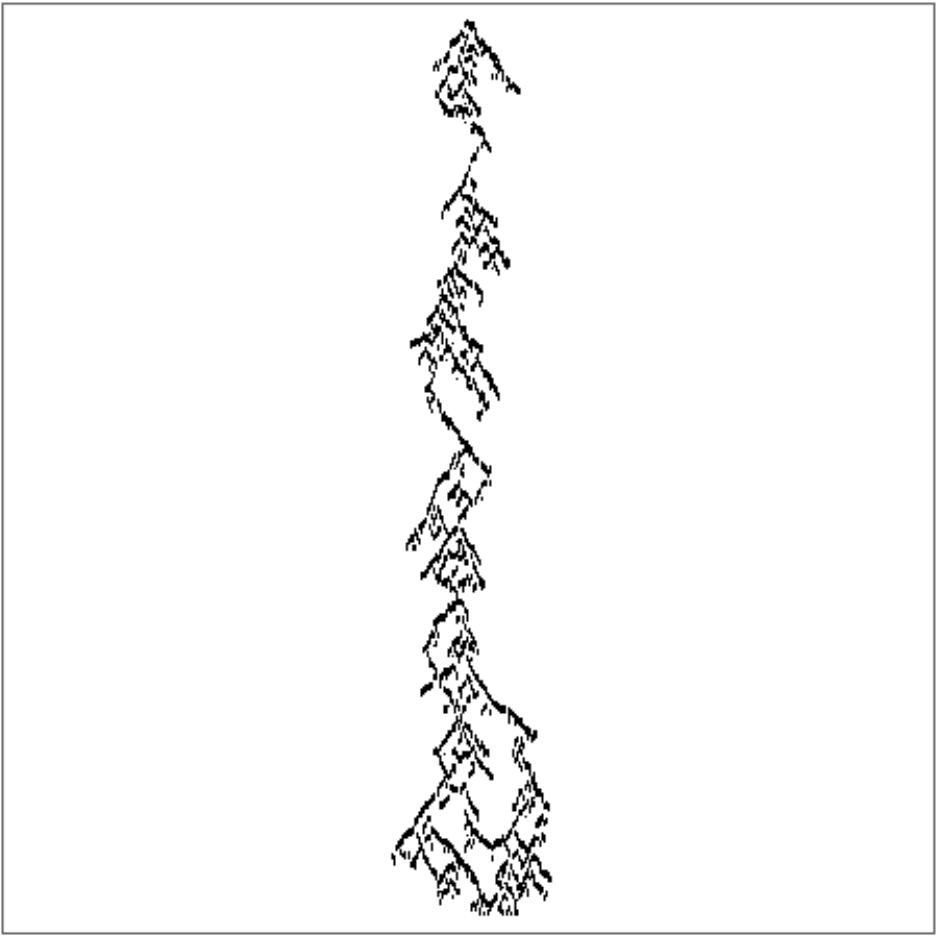}
 \includegraphics[width=50mm,height=50mm]{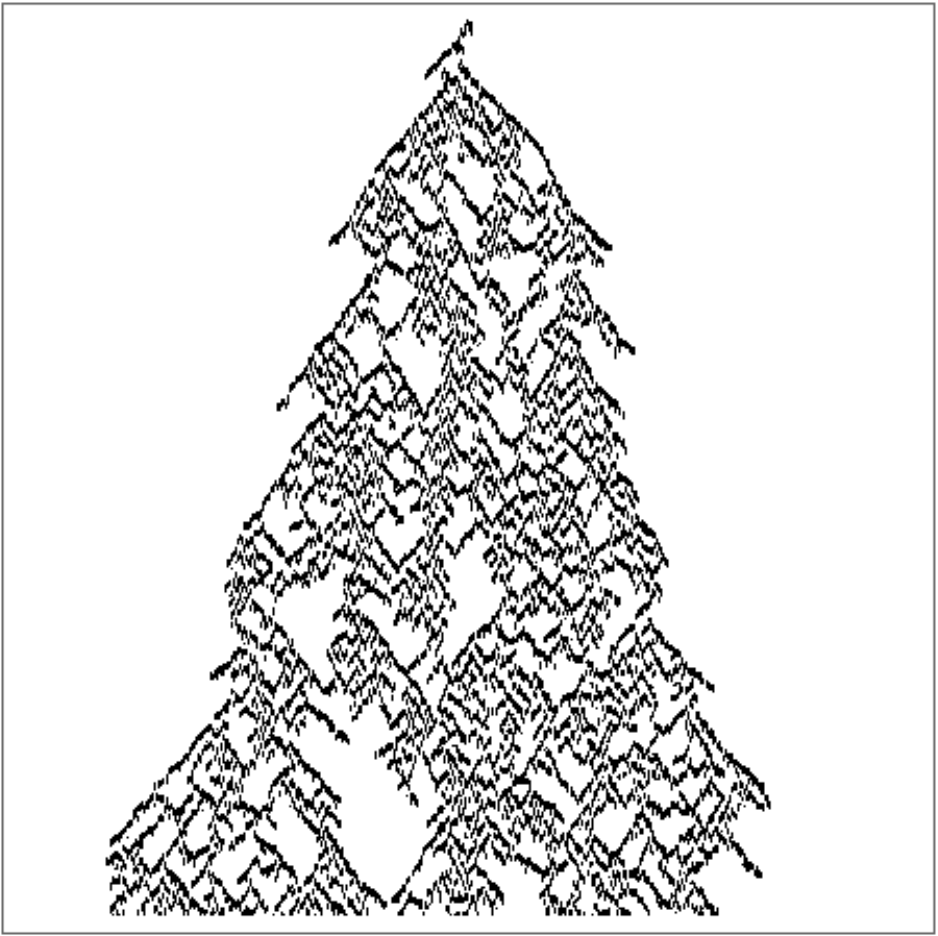}
\caption{Typical space-time evolutions for $\lambda=0.090$ (right) and $\lambda=0.094$ (left). Simulation starts from a single active particle (black) and time increases downward.}
 \end{figure}

To confirm if the phase transition in this case is in the DP universality class, we are going to calculate some values of the critical exponents of this model, and before that we will determine the value of critical point $\lambda_{c1}$. In Fig. 4 we plot the average steady state of density of active particles $\rho_s$ at various values of the infection probability $\lambda$. In the simulation, we use a lattice of size $N=10^4$ averaged over $200$ realizations after discarding the first $10^4$ time steps. Fig. 4 shows clearly that, the system crosses from the absorbing phase to the active phase at specific value of the parameter $\lambda$. For the best estimations, the value of the critical point seems to converge to $\lambda_{c1}=0.906\pm 0.004$. Using this result we can determine one of the critical exponents related to this model where, it is known that, for the continuous phase transitions and as the control parameter $\lambda$ approaches the critical point $\lambda_c$, the stationary value of the order parameter $\rho_s$ vanishes asymptotically according to a power law as follows \cite{hen,hin,gez}:
\begin{eqnarray}
\rho_s\sim (\lambda-\lambda_c)^{\beta}
\end{eqnarray}

Inset of Fig. 4 shows the logarithmic plot of $\rho_s$ as function of the distance from the critical point $(\lambda-\lambda_{c1})$, which shows clearly the power law behavior.
Estimated value of the critical exponent $\beta$ from the inset of Fig. 4 gives us $\beta=0.281\pm 0.005$ which consists very well with the value of $\beta=0.276$ for the $(1+1)$ DP universality class \cite{hen,hin,gez}.

\begin{figure}[htb]
 \includegraphics[width=100mm,height=80mm]{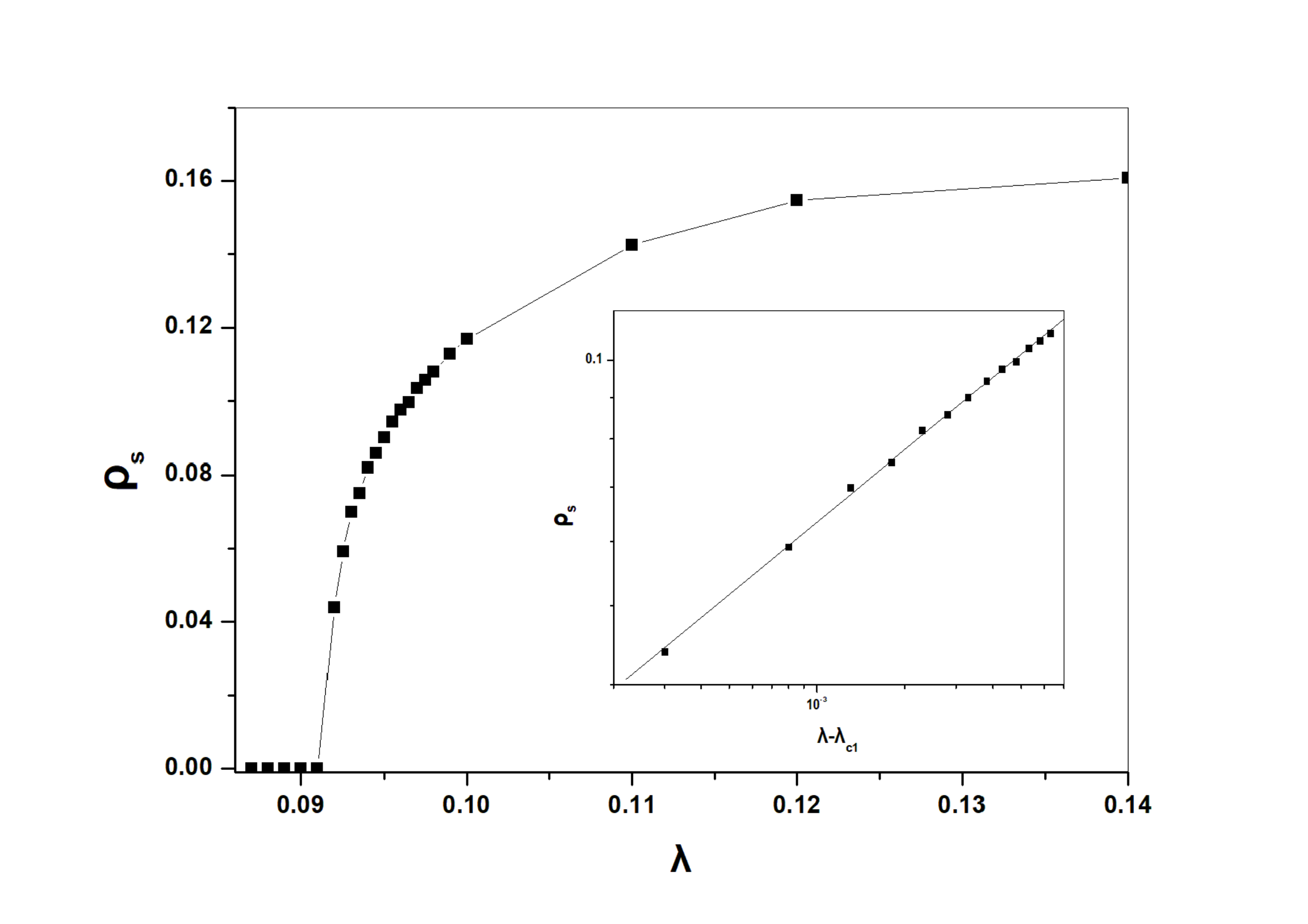}
\caption{The steady state of density of particles at various values of the $\lambda$ for the same parameters in Fig. 2. The inset is Log-log plot of the $\rho_s$ as function of the distance to the critical point.}
 \end{figure}

To extract a further critical exponent, we perform time dependent Monte Carlo simulations of this model starting from a fully occupied lattice. As we know for continuous phase transitions, at the critical point $\lambda_c$, the time evolution of the order parameter $\rho(t)$ decays asymptotically according to the following power law \cite{hen,hin,gez}
\begin{eqnarray}
\rho(t)\sim t^{-\delta}
\end{eqnarray} 
Where $\delta$ is the critical exponent which equal to $0.159464(6)$ for DP universality class in the $1+1$ dimension \cite{hin}. 

Fig. 5 shows the density of active particle $\rho(t)$ as function of time on a logarithmic scale. At the critical point the system clearly shows a power law decay of the active particles. For the best fitting, the value of the critical exponent we find to be $\delta=0.159\pm 0.005$ which is again consistent with the value of the critical exponent for DP universality class in the $1+1$ dimension.  
\begin{figure}[htb]
 \includegraphics[width=100mm,height=80mm]{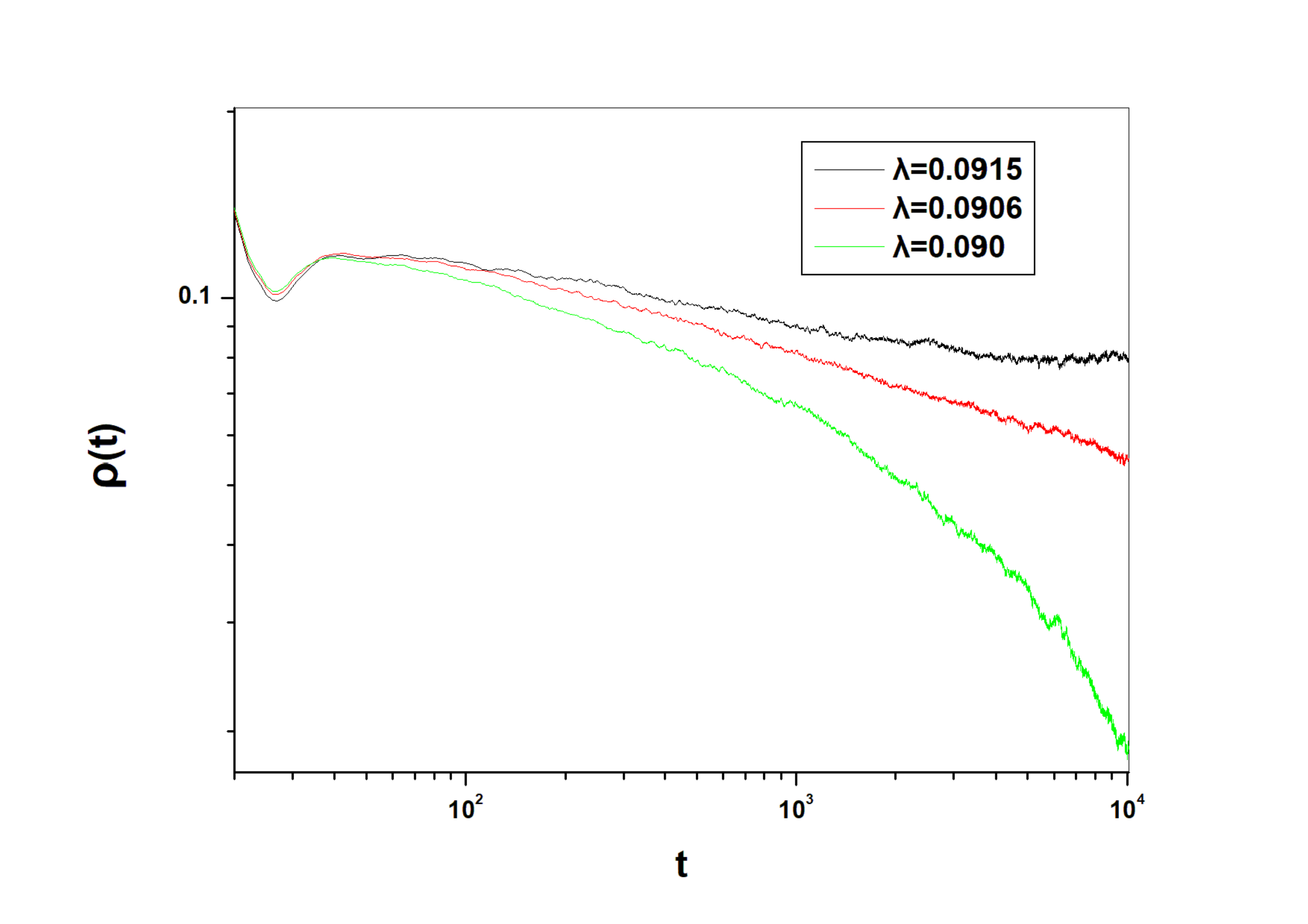}
\caption{The density of active particles as function of $\lambda$ when $N=10^4$, $k=3$, $\tau_I=7$ and $\tau_R=9$.}
 \end{figure}

Hence we can confirm that, for the case when $\tau_I<\tau_R$ the phase transition from the absorbing state to active state for SIRS model at the first critical point $\lambda_{c1}$ is of kind DP universality class in $(1+1)$ dimension. Here we should mention to that, close to $\lambda_{c1}$, the values of $\lambda$ are low enough in which the stable absorbing state $S$ is dominated state on the system i. e., it is impossible for the system to approach the un-stationary absorbing state $I$ in this case. Therefore the accessible absorbing state for the system close to the first critical point only the state $S$, consequently in this case the system satisfies Janssen Grassberger criterion expect this model is multi-component system.

\section{Phase Transition at the second critical point $\lambda_{c2}$}
As we increase the value of $\lambda$ toward the second critical point $\lambda_{c2}$ we observe that, at a specific value of $\lambda$ (which is $\lambda>0.15$ for the system of size $N=10^4$ other parameters as same as the parameters we have used in the previous section) the steady state of average density of infected particles $\rho(t)$ becomes strongly dependent on its initial values $\rho(0)$ as the Fig. 6 shows. In Fig. 6 we plot the average values of $\rho(t)$ as function of time for various values of initial conditions $\rho(0)$. Fig. 6b, shows three plots of $\rho(t)$ as function of time for the case when $\lambda=0.16$, in which we start simulations from different initial values $\rho(0)$. The figure shows clearly that, for the initial values $\rho(0)=0.1$ and $\rho(0)=0.3$ the system attains the same steady state, however when $\rho(0)=0.6$ the system saturates at a different steady state. Fig. 6c shows the situation for $\lambda=0.20$ with initial conditions same as in Fig. 6b. In this case the system dependence on its initial values becomes more apparent, where the three plots approach different steady states for the three values of initial conditions. For $\lambda=0.12$, Fig. 6a is given for comparison, where in this case the curves reach same steady state and the system does not depend on its initial conditions. Here we should mention to that, in Monte Carlo simulations of Fig. 6, we take the average of $\rho(t)$ over all configurations those survive or not. Another point we mention here is that, the sensitivity of the steady state of the order parameter to initial conditions in the model of disease spreading has been observed in minimal vaccination-epidemic model \cite{pir}.  
 
\begin{figure}[htb]
 \includegraphics[width=50mm,height=50mm]{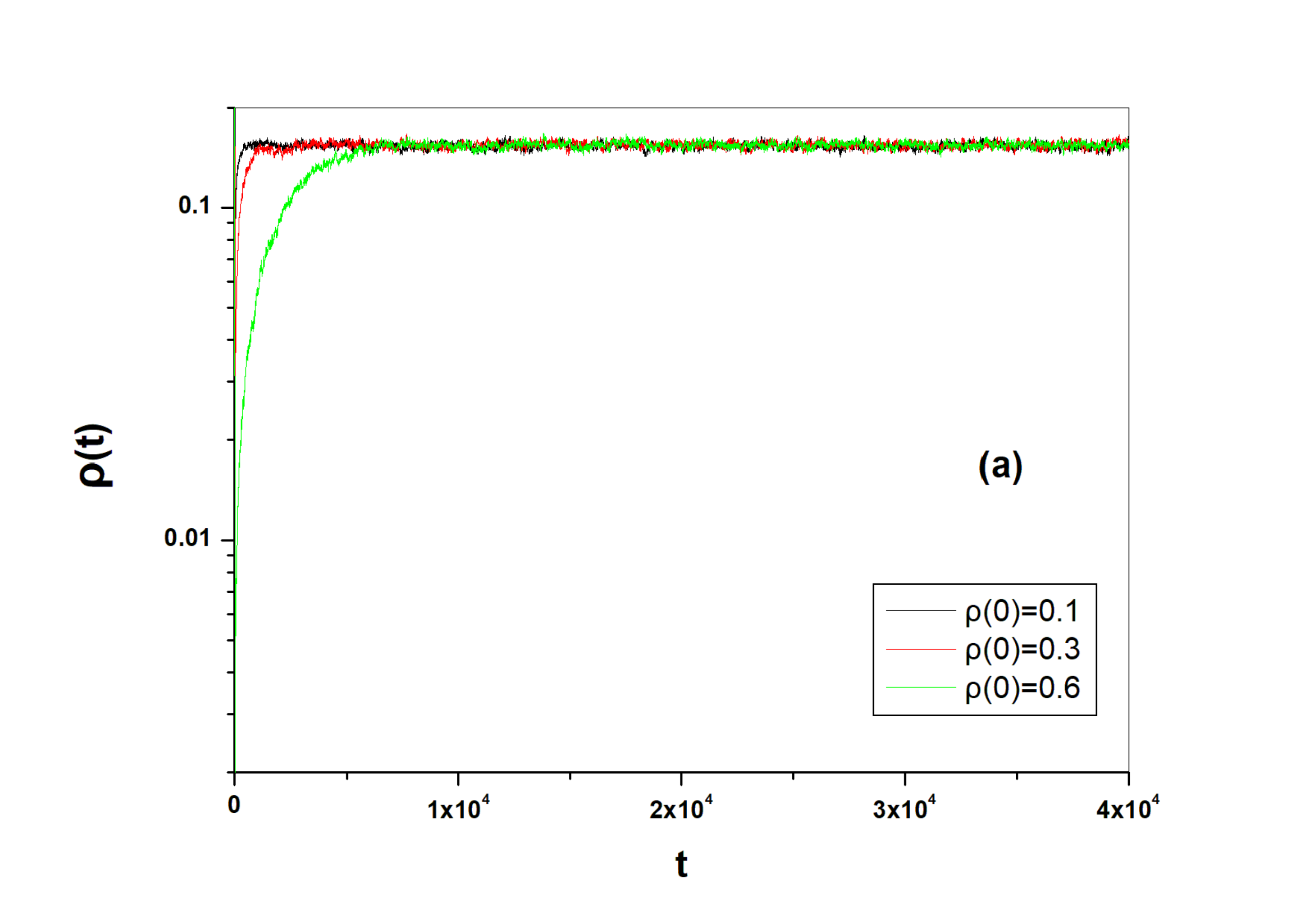}
 \includegraphics[width=50mm,height=50mm]{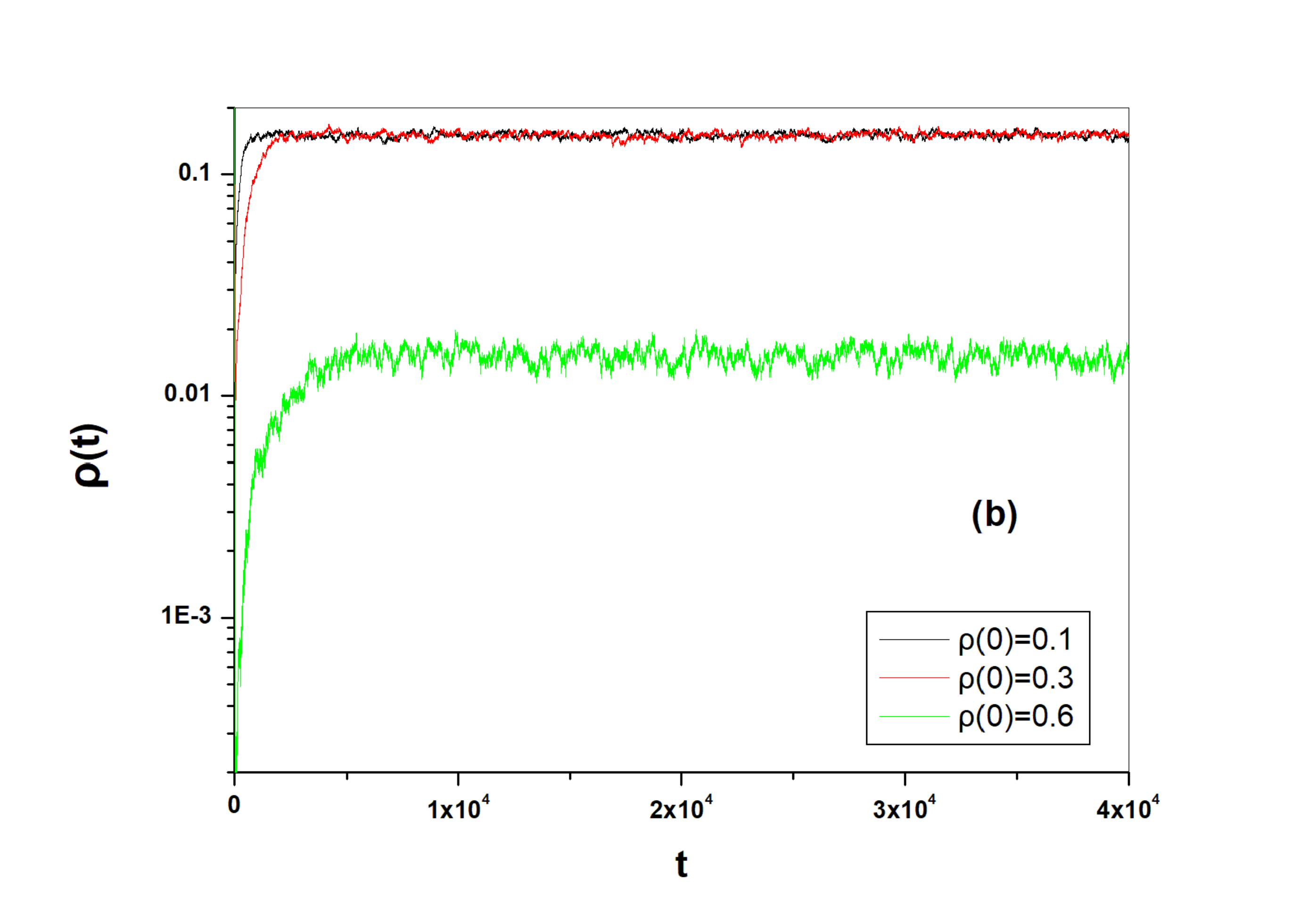}
 \includegraphics[width=50mm,height=50mm]{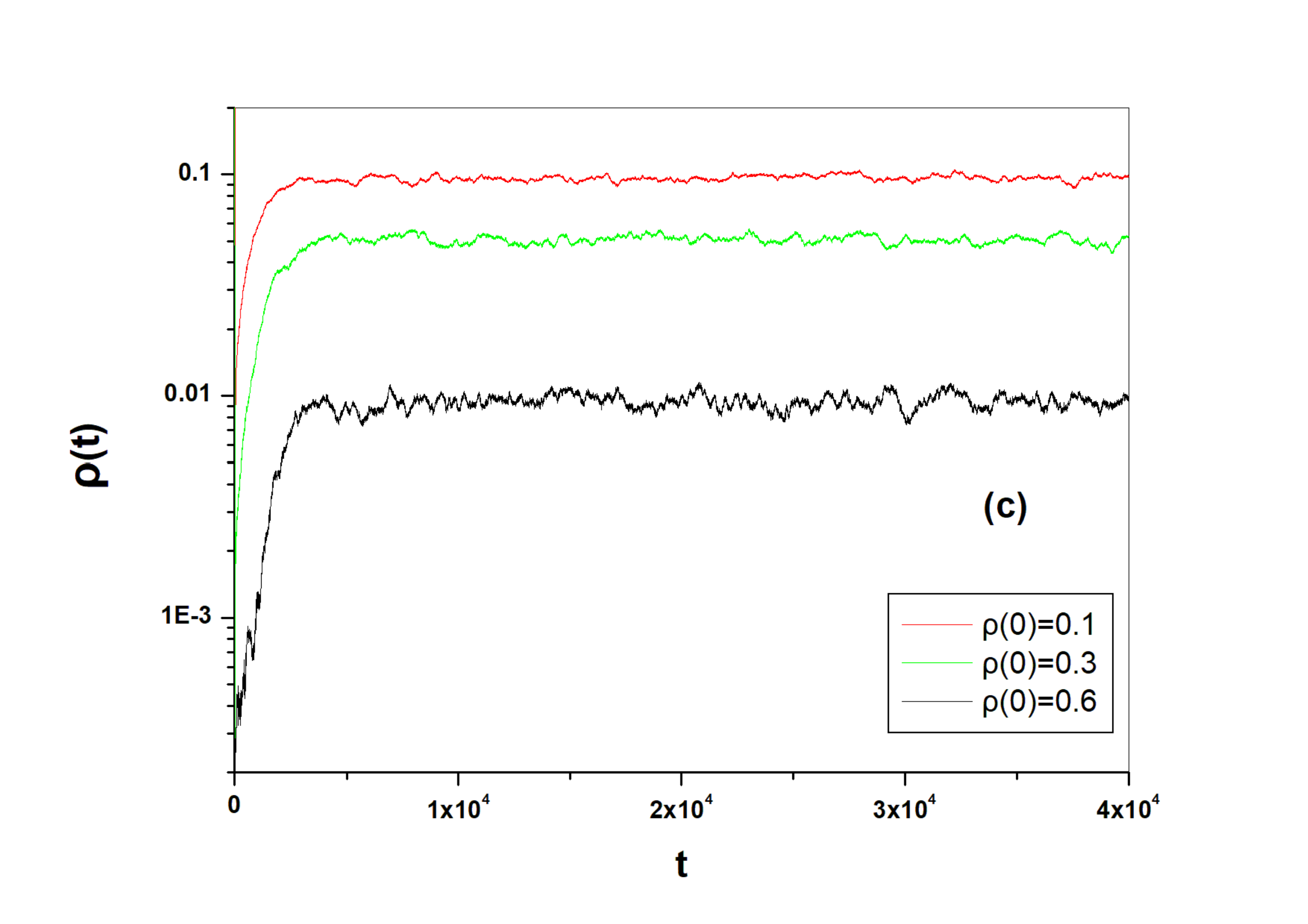}
\caption{The time evolution of average value of density of particles as function of time at different values of initial conditions for $N=10^4$, $k=3$, $\tau_I=7$ and $\tau_R=9$: a) for $\lambda=0.12$ b) for $\lambda=0.16$ c) for $\lambda=0.20$. For each curve we averaged over $200$ realizations.}
 \end{figure}
  
By careful inspection of the time evolution of $\rho(t)$ we observe that, there are some configurations quickly become trapped in absorbing state during a time which is not longer than $\tau_I+\tau_R$ (in our case it is $17$ time steps). This time is on the average is exactly the time need it the particles to go through one infection cycle. Whereas some other configurations survive for a longer time. We also observe that, for fixed values of $\rho(0)$ the density of trapped configurations increases as the value of $\lambda$ becomes higher. In contrast increasing in the values of $\rho(0)$ causes increasing in the density of trapped configurations for fixed values of $\lambda$. Fig. 7 shows the density of trapped configurations (DTCO) as function of density of initial active particles $\rho(0)$ for the value of $\lambda=0.16$. In this figure, the trapped configurations are those configurations reach the absorbing state during the time which is less than or equal $17$ time steps. Figure shows clearly that, whenever the initial density $\rho(0)$ of active particles is $\rho(0)>0.4$, all the configurations reach the absorbing state. However, when the values of $\rho(0)<0.4$, we can see increasing in the number of surviving configurations as the value of $\rho(0)$ decreases.

\begin{figure}[htb]
 \includegraphics[width=80mm,height=80mm]{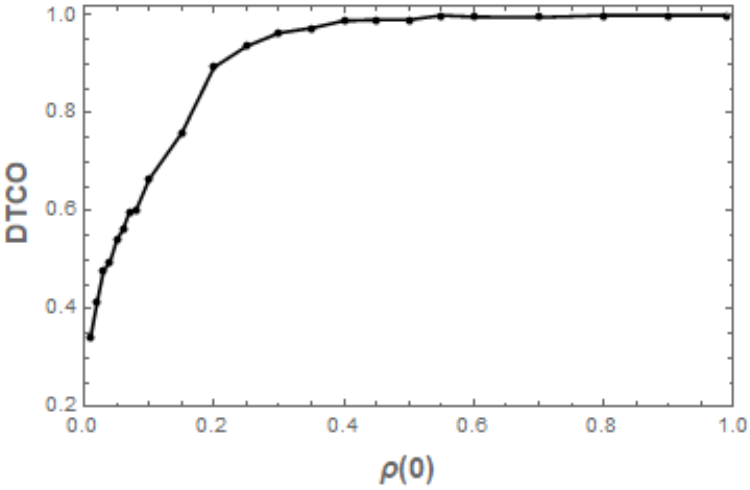}
\caption{Density of trapped configurations (DTCO) as function of time at different values of initial values conditions when $N=10^4$, $k=3$, $\tau_I=7$, $\tau_R=9$, and $\lambda=0.16$, each point in the figure averaged over $2000$ configurations.}
 \end{figure}
 
We can deduce from the previous results that, the chance for the system to approach the state where all the particles in the network are infected (absorbing state $I$) becomes possible as the value of $\lambda$ goes toward the higher values. This possibility enhances as the density of initial values are increased. We can understand the relation between the values of infection probability $\lambda$ and the initial values density $\rho(0)$ of this model from the mean field approximation (see Ref. \cite{gon}). Whereas the increasing on the number of $I$ particles at the next time will proportional to the number of particles of kind $I$ and $S$ at this time, i. e.  $I(t+1)\propto I(t)+\lambda S(t) I(t)$, then, if the value of $\lambda S(I) I(t)$  becomes high enough such that $I(t+1)=N$ the system will reach the un-stationary absorbing state $I$. Therefore, we can say that, for high values of $\lambda$ un-stationary absorbing state becomes the dominating state in system.

Because of that dependence for the system on its initial values conditions near $\lambda_{c2}$, we have faced difficulty in determining the kind of the phase transition or even accurately determine the critical point close to $\lambda_{c2}$. We should mention also to that, we could not get any kind of power law behavior near the expected value of the critical point using the time dependent dynamics starting from a fully occupied lattice or form a single active seed. However, and according to Refs. \cite{dic1,odo1}, the initial values dependence is an indicator of a discontinuous phase transition. Therefore, for better understanding about the mechanism of the phase transition in this case, we perform the order parameter quasistationary probability distribution for this model. As claimed by Refs. \cite{oli3,win,mai,oli4,san,dic2} the order parameter quasistationary probability distribution is bimodal in the neighborhood of a discontinuous phase transition in contrast to a continuous phase transition where there will be only a single pick. Fig. 8 shows the results of our Monte Carlo simulations for the cell-occupancy histogram distribution (P) of our model for cells of $100$ sites at the center of lattice of $N=10^3$ particles, at a various values of $\lambda$. The variable $n$ in Fig. 8 is the number of active particles. The quasistationary distribution is clearly bimodal distribution. This result enhances assumption the discontinuous phase transition of this model at $\lambda_{c2}$. For comparison the inset of Fig. 8 shows the cell-occupancy histogram distribution in the vicinity of $\lambda_{c1}$ ($\lambda=0.1$), where the system undergoes a continuous phase transition. It is clear in this case there is only a single pick.

\begin{figure}[htb]
\includegraphics[width=80mm,height=80mm]{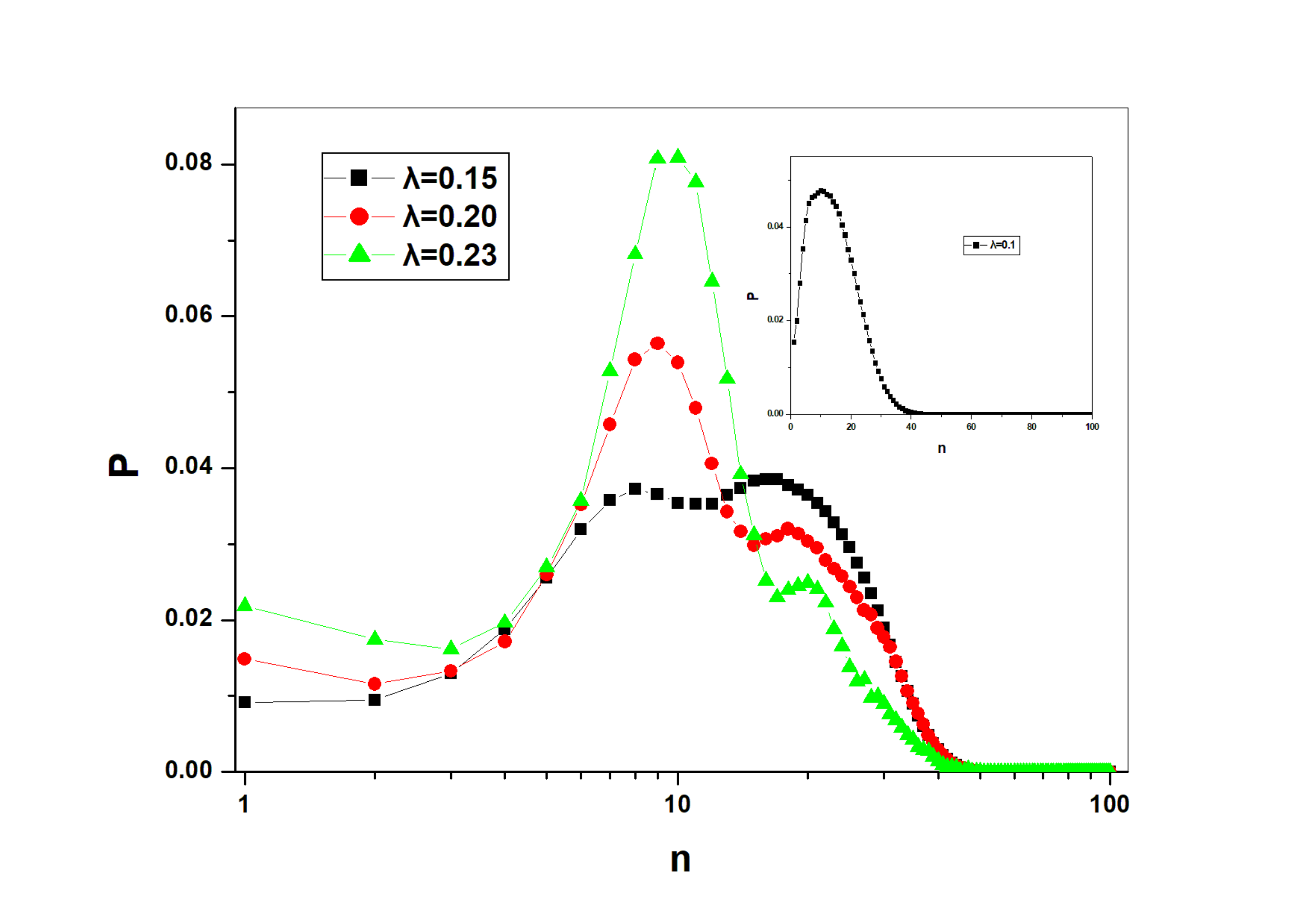}
\caption{Cell-occupancy histogram distribution (P) for cells of 100 sites, when $\lambda=0.15$, $\lambda=0.20$ and $\lambda=0.23$, $N=10^3$, $k=3$, $\tau_I=7$ and $\tau_R=9$. Inset: Cell-occupancy histogram distribution for the same parameters expect $\lambda=0.1$.}
 \end{figure}

The results we find previously suggest strongly that, the phase transition at the second critical point $\lambda_{c2}$ is discontinuous. However, for a more clarification we study the quasistationary behavior of this system beside the critical point $\lambda_{c2}$. For that we achieve a finite-size analysis, which is a more reliable procedure as recently proposed in Refs. \cite{oli2,net}. According to that procedure, the difference between the pseudotransition point $\lambda_N$ (where $N$ denotes the system volume) and the transition point $\lambda_{c2}$ scales with $N^{-1}$ according to the relation $\lambda_N=\lambda_{c2}+aN^{-1}$.
Therefore, in order to determine accurately the value of $\lambda_N$ we use the system order parameter variance $\chi= N (\left\langle \rho^2\right\rangle-\left\langle \rho\right\rangle^2)$. This quantity has been proven to has a peak at the value of pseudotransition point $\lambda_N$ \cite{oli2,net}. We restrict our simulation here, only on surviving configurations. Fig. 9a shows the order parameter variance $\chi$ as function of $\lambda$. Whereas Fig. 9b shows how the values of pseudotransition point $\lambda_L$ scale with the values of system size $N^{-1}$. Extrapolation of $N\rightarrow \infty$ yields the critical point for this model to be $\lambda_{c2}=0.24\pm 0.01$.
\begin{figure}[htb]
 \includegraphics[width=80mm,height=80mm]{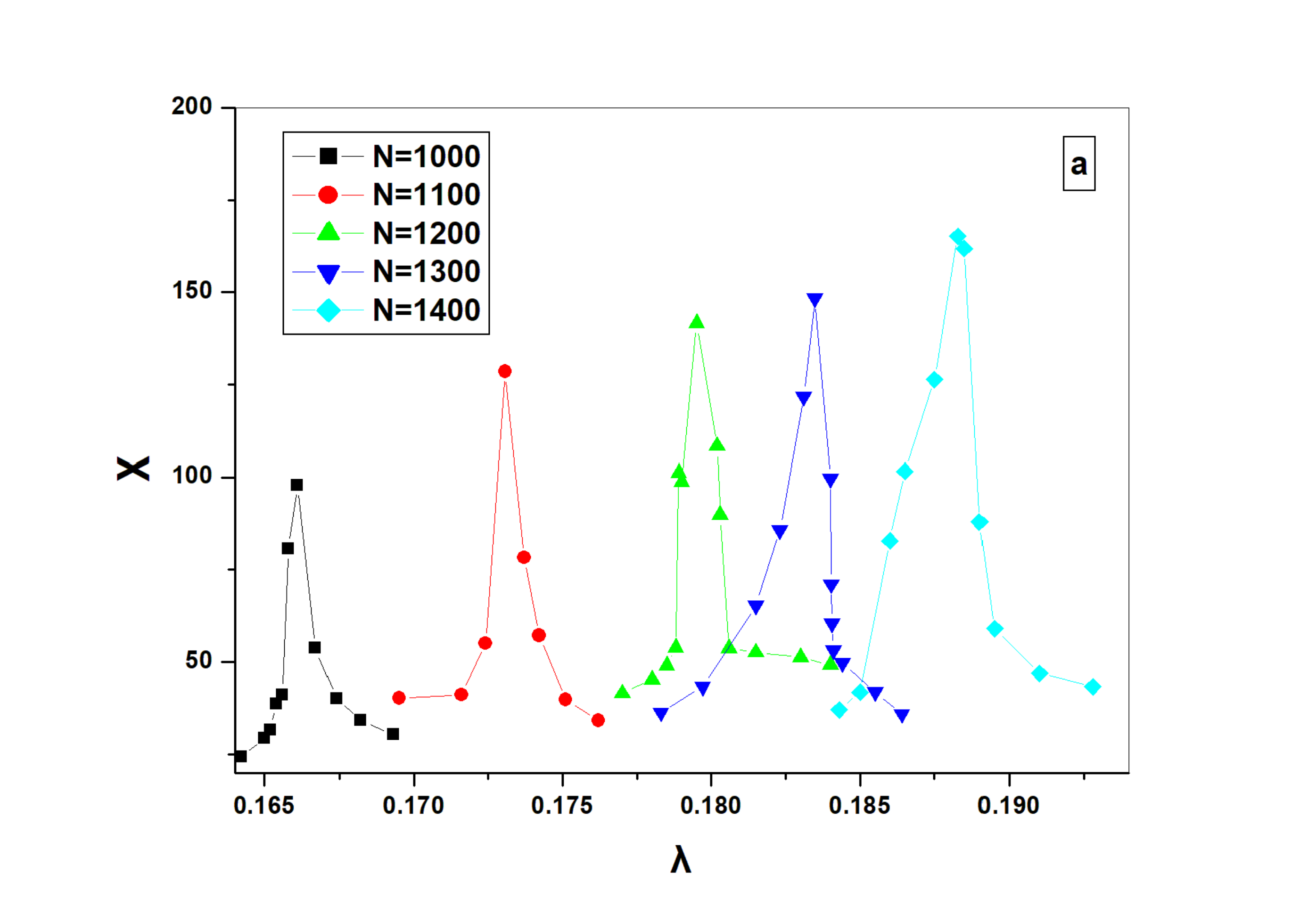}
 \includegraphics[width=80mm,height=80mm]{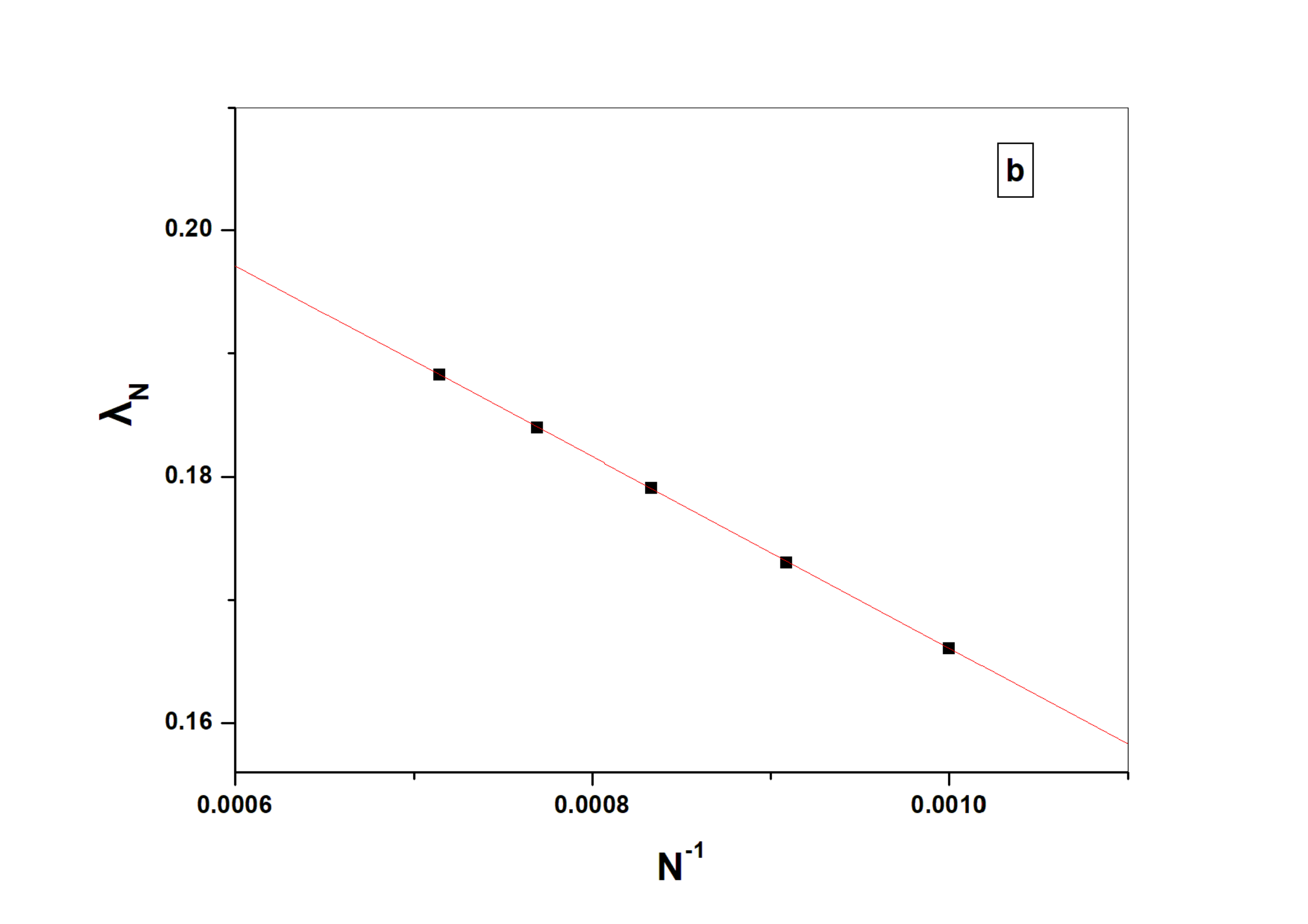}
\caption{(a) Order parameter variance $\chi$ versus the parameter $\lambda$. (b) Value of $\lambda_N$ for which $\chi$ is maximum vs $N^{-1}$}
 \end{figure}

Final point we have studied for this model is the space-time evolution of the active particles close to $\lambda_{c2}$. Fig. 10 shows the infected particles with red color during the time of $10^3$ time steps for a system of $N=10^3$ particles at the value of $\lambda=0.231$ and $\lambda=0.241$. In both figures simulation starts at $t=0$ with all particles are in the state $S$ expect for a one active particle $I$ at the center of lattice. Figures clearly show that, the spreading of active particles are compact in a behavior remind us the spreading behavior of CDP models. However, the difference here is that, the particles have a finite time to stay in the active phase. The coexistence of small compact isolated islands of active particles with high regions of inactive ones is again indeed consistent with a discontinuous phase transition.
\begin{figure}[htb]
 \includegraphics[width=50mm,height=50mm]{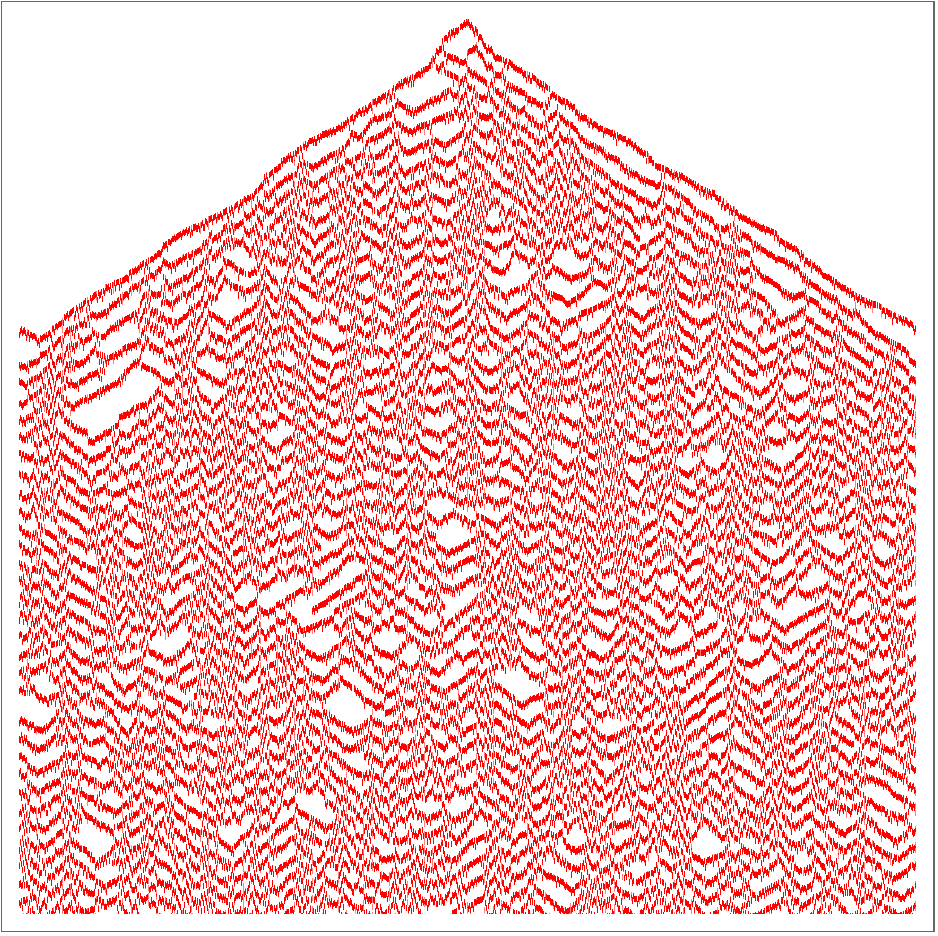}
 \includegraphics[width=50mm,height=50mm]{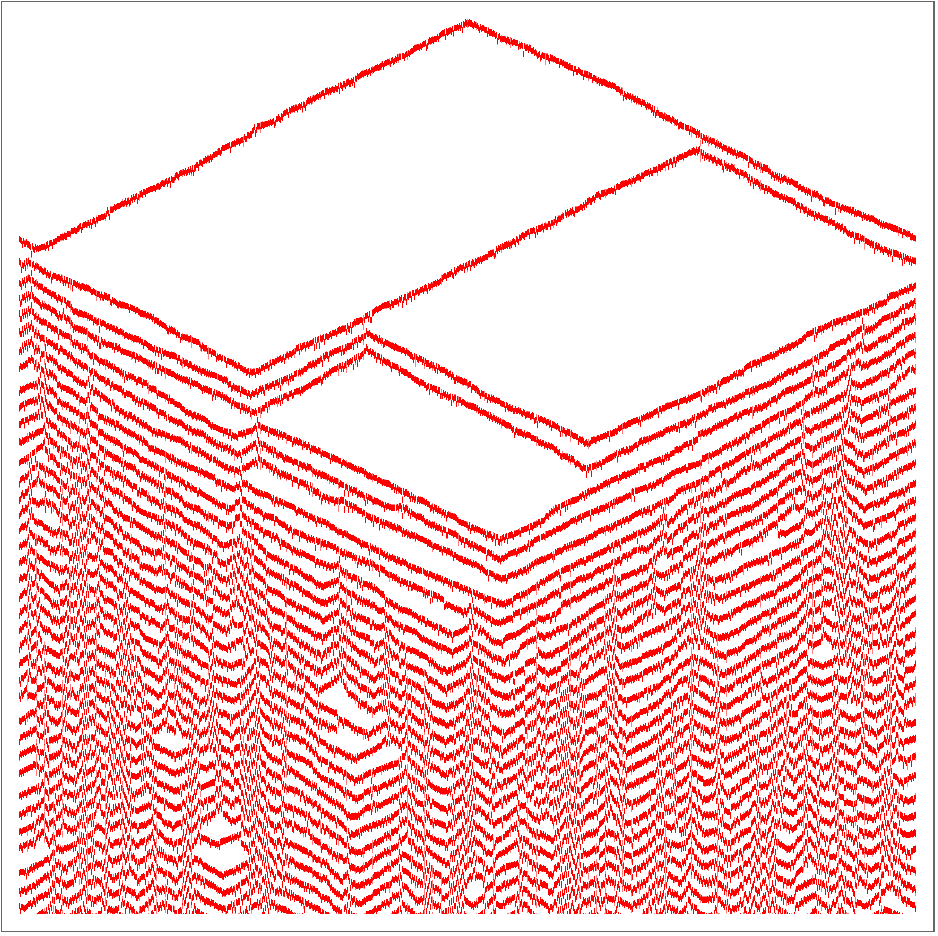}
\caption{Space-time evolution of SIRS model from a single seed (red) when $\lambda=0.231$ (right) and $\lambda=0.241$ (left). Other parameters are $N=10^3$, $k=3$, $\tau_I=7$ and $\tau_R=9$.}
 \end{figure}
  
Finally we mention to that, the models of disease spreading such as a minimal vaccination-epidemic model and Susceptible- Infected-Susceptible (SIS) model have been also found to show either a continuous or a discontinuous active to absorbing phase transition \cite{pir,ma,pi}. Additionally, we can deduce some similarities between the phase transition in this model and the phase transition of ZGB model on two dimensional lattice where both models have two critical thresholds. In both models the first critical threshold is corresponding to the continuous DP class and the second critical threshold is corresponding to the discontinuous phase transition. ZGB model has two absorbing states, the first one is at small values of adsorption rate (beside the first critical point) and the second one (beside second critical point) is at high values of adsorption rate \cite{zif}. SIRS also has a one absorbing state at a low infection rate (beside the first critical point) and unstable absorbing state at a high infection rate (beside second critical point).

\section{Conclusions}
In the summary, we have studied the phase transition from the absorbing phase to active phase for the model of infection spearing SIRS on the one dimensional network. This model has been found to has a two critical points where the infection survives in between those critical points and dies out outside of them. The two critical points correspond to low infection rate and high infection rate. Using Monte Carlo simulations we have found that, whereas the the phase transition at the first critical point is of kind the DP universal class, the phase transition at the second critical point is of kind first order phase transition. In this manner, the presence of continuous and discontinuous phase transitions has been also confirmed in the models of disease spreading suchas a minimal vaccination-epidemic model and SIS model \cite{pir,ma,pi}. We can also compare the phase transition in this model with the phase transition in ZGB model. Both models have two critical points, where the phase transition at the first critical point is of kind DP class and the phase transition at the second critical point is discontinuous. However, we should mention here to that, the system we have studied here is a one dimensional system whereas ZGB is a two dimensional system.


 \end{document}